\documentclass[aps,preprint,showpacs,preprintnumbers,amsmath,amssymb]{revtex4}

\usepackage{graphicx}
\usepackage{dcolumn}
\usepackage{bm}

\newcommand{\lam}{\Lambda}

\newcommand{\lop}{\mathcal{L}}
\newcommand{\dop}{\mathcal{D}}
\newcommand{\tlop}{\widetilde{\mathcal{L}}}
\newcommand{\bx}{\overline{x}}
\newcommand{\tphi}{\widetilde{\phi}}
\newcommand{\ta}{\widetilde{a}}

\begin{document}

\title{Eigenvalue Estimation of Differential Operators with a Quantum Algorithm}

\author{Thomas Szkopek}
 \email{szkopek@ee.ucla.edu}
\affiliation{Department of Electrical Engineering, University of %
California Los Angeles, Los Angeles, California 90095}
\author{Daniel S. Abrams}
\affiliation{Luminescent Technologies, Inc., Mountain View, California 94041}
\author{Vwani Roychowdhury}
\author{Eli Yablonovitch}
\affiliation{Department of Electrical Engineering, University of %
California Los Angeles, Los Angeles, California 90095}

\date{\today}

\begin{abstract}
We demonstrate how linear differential operators could be emulated by a quantum processor, should
one ever be built, using the Abrams-Lloyd algorithm. Given a linear differential operator of order
$2S$, acting on functions $\psi(x_1,x_2,\ldots,x_D)$ with $D$ arguments, the computational cost
required to estimate a low order eigenvalue to accuracy $\Theta(1/N^2)$ is
$\Theta((2(S+1)(1+1/\nu)+D)\log N)$ qubits and $O(N^{2(S+1)(1+1/\nu)}\log^cN^D)$ gate operations,
where $N$ is the number of points to which each argument is discretized, $\nu$ and $c$ are
implementation dependent constants of $O(1)$. Optimal classical methods require $\Theta(N^D)$ bits
and $\Omega(N^D)$ gate operations to perform the same eigenvalue estimation. The Abrams-Lloyd
algorithm thereby leads to exponential reduction in memory and polynomial reduction in gate
operations, provided the domain has sufficiently large dimension $D > 2(S+1)(1+1/\nu)$. In the case
of Schr\"{o}dinger's equation, ground state energy estimation of two or more particles can in
principle be performed with fewer quantum mechanical gates than classical gates.
\end{abstract}


\maketitle

An early motivation for research in quantum information processing has been the simulation of
quantum mechanical systems \cite{feyn}. The Abrams-Lloyd algorithm \cite{lloyd96,abrams97,abrams99}
is an instance of quantum mechanical simulation (followed by variations \cite{boghosian97},
\cite{zalka98}, \cite{meyer01}). We describe in this paper the application of the Abrams-Lloyd
algorithm to estimating low order eigenvalues of linear partial differential equations with
homogeneous boundary conditions (more precisely, Hermitian boundary value problems). The
significance of our analysis is two fold. First, we generalize the Abrams-Lloyd algorithm to
boundary value problems other than Schr\"{o}dinger's equation, which may find application to
classical problems. Secondly, we quantify computational cost and determine under what conditions we
may expect the Abrams-Lloyd algorithm to give a reduction in computational work compared to optimal
classical techniques in order to achieve the same eigenvalue accuracy.

Very briefly, the Abrams-Lloyd algorithm as originally envisaged for the many-body Schr\"{o}dinger
equation is structured as follows. An initial estimate $|\psi(0)\rangle$ of the target eigenstate
is loaded into a multiple qubit register. Controlled application of a unitary operation, chosen to
correspond to the time evolution operator $\exp(-i\mathcal{H}\tau)$ of the many-body Hamiltonian
$\mathcal{H}$ under study for time step $\tau$, allows one to generate a sequence of time evolved
states originating from the initial guess,
$\{|\psi(0)\rangle,|\psi(\tau)\rangle,|\psi(2\tau)\rangle,\ldots\}$. A spectral analysis of the
sequence of time evolved states recovers the frequency (energy) of the target eigenstate (provided
the initial guess was ``close enough''). The Abrams-Lloyd algorithm is akin to a stroboscope for
quantum states evolving under a many-body Hamiltonian. If the total time of evolution is
sufficiently long, while the individual time steps are sufficiently small, a high frequency
(energy) resolution can be achieved. Following the determination of the eigenvalue, the
corresponding eigenstate remains in the qubit register. Although the full amplitude description of
an eigenstate is inaccessible, some information about the state can be extracted to a precision
ultimately limited by the number of qubits used to represent the eigenstate (so for instance, one
can test symmetries of the eigenstate).

The algorithm can be extended to more general partial differential equations rather easily. So long
as the boundary value problem is Hermitian, we can map our mathematical problem to a fictional
quantum system and apply the algorithm without change. The partial differential operator, $\lop$,
will correspond to a (possibly) fictional Hamiltonian $\mathcal{H}$, and an initial guess $\psi(0)$
will correspond to an initial wavefunction $|\psi(0)\rangle$. In other words, quantum mechanical
amplitudes represent function values. Less obviously, the sequence of time evolved states,
$\{|\psi(0)\rangle,|\psi(\tau)\rangle,|\psi(2\tau)\rangle,\ldots\}$\, has a mathematical analogue
of great use in classical matrix eigenvalue analysis, known as the Krylov subspace:
$\mathrm{span}\{\psi(0),\exp(-i\lop \tau)\psi(0),\exp(-i2\lop \tau)\psi(0),\ldots \}$. The subspace
is generated by repeated application of $\exp(-i\lop \tau)$, although in classical techniques one
more typically uses rational functions of $\lop$. Here, $\tau$ no longer has the physical meaning
of time. Rather, $\tau$ sets the scale for how much phase is applied per application of
$\exp(-i\lop \tau)$. As in the quantum simulation, a large total phase applied one small phase step
at a time allows a high resolution estimate of eigenvalues. We quantify these notions now.

The computational cost of the Abrams-Lloyd algorithm for a specified eigenvalue accuracy is limited
as a consequence of three sources of error, expressed here in the language of quantum mechanical
simulation:
\newcounter{lcount}
\begin{list}{\Roman{lcount}}{\usecounter{lcount}}
\item \emph{truncation error}: Discretization is necessary for a computational model based on
qubits. However, discretization of the continuous problem to $N$ points per coordinate results in
$\Theta(1/N^2)$ relative error in low order energy eigenvalues due to truncation of high spatial
frequency contributions. The choice of $N$ must be made appropriate to the accuracy that is desired. \\
\item \emph{splitting error}: The full many-body evolution $\exp(-i\mathcal{H}\tau)$ over time step
$\tau$ can be implemented with universal gates by splitting the full evolution into a sequence of
efficiently implementable unitaries $\exp(-i\mathcal{H}^{k}\tau)$, where $\mathcal{H} = \sum_k
\mathcal{H}^{k}$. The approximation results in an absolute eigenvalue error $O(\left\| \mathcal{H}
\right\|_2^{\nu+1}\tau^{\nu})$, where $\|\mathcal{H}\|_2$ is the maximum eigenvalue of the
discretized Hamiltonian $\mathcal{H}$ and $\nu$ is a constant of $O(1)$ determined by the precise
sequence of local operators chosen. Splitting error requires us to use small time steps $\tau$.
\\
\item \emph{frequency resolution}: A quantum Fourier transform, like any discrete Fourier
transform, can resolve absolute phase to accuracy at best $\pm \pi$. For a sequence of $M$ samples,
the relative error in an energy eigenvalue $E$ will be $\pm \pi/(M E \tau)$. Frequency resolution
requires us to simulate over a large total time $M\tau$.
\\
\end{list}
The optimal way to balance these errors is as follows. Since we are interested in the continuous
problem, we first choose a discretization of $N$ points per sample so that the discrete problem
eigenvalue approximates the continuous eigenvalue problem to some desired accuracy $\Theta(1/N^2)$.
We wish to solve the discretized problem to an accuracy determined by the truncation error limit;
solving the discrete problem to greater accuracy leads to wasted effort since we are interested in
the continuous problem, while solving the discrete problem to lesser accuracy implies we have
wasted effort by choosing too many discrete points $N$ per coordinate. We can thereby determine the
maximum time step $\tau$ to keep splitting error no greater than truncation error. Next, we can
determine the number of time steps $M$ required to resolve eigenvalues with the quantum Fourier
transform at the truncation error limit. In the case of Hermitian boundary value problems, We show
in this paper that the resulting computational cost is $\Theta(D \log N)$ qubits and
$O(N^{2(S+1)(1+1/\nu)}\log^cN)$ gate operations, where $2S$ is the differential order of $\lop$ and
$c$ is a constant $O(1)$. This can be compared with the optimal classical cost of $\Theta(N^D)$
bits and $\Omega(N^D)$ gate operations. Near optimal classical methods approaching these costs do
in fact exist \footnote{A near optimal classical method can be constructed using a combination of
Krylov subspace iteration, matrix preconditioning and multigrid solutions, as in \cite{brandt83}.
See \cite{demmel},\cite{saad} for a sampling of the vast array of classical numerical techniques
available \label{note:classicaltechniques}}.

We emphasize that in our analysis, we take a constructive approach wherein we account for all the
logical operations required to implement the algorithm without recourse to oracles that may or may
not have physically efficient implementations. This is in contrast to previous work including the
simulation of spin glass physics \cite{lidar97}, and Sturm-Liouville problems
(\cite{papageorgiou05} and references therin). As stated, our motivation is to compare the
computational cost of eigenvalue estimation by the Abrams-Lloyd algorithm and optimal classical
methods.

Our paper is organized as follows. In section \ref{sec:reduction}, we introduce the one dimensional
eigenvalue problem, which will serve as a useful example with which the principles of the algorithm
can be illustrated. We derive the truncation error for low order eigenvalues in a way suitable for
extension to higher dimensional problems. The algorithm itself is described in section
\ref{sec:quantalg}, followed by an analysis of computational cost as it is applied to the one
dimensional problem in section \ref{sec:compcost}. A circuit suitable for a 2nd order differential
equation is given as a concrete example. Generalization of the algorithm to higher dimensional
problems is given in section \ref{sec:generalization} along with an analysis of computational cost,
where we show a reduction in computational work polynomial in $N$ over classical techniques.
Concluding remarks about the computational efficiency of the Abrams-Lloyd algorithm are given in
section \ref{sec:conclusion}.

\section{One-Dimensional Problem}

\label{sec:reduction}

To illustrate the essential features of eigenvalue estimation of
differential operators, it's instructive to consider a Hermitian
one-dimensional problem, which we introduce here in some detail. The
primary result of this section is a derivation of truncation error
in low order eigenvalues as a function of discretization. Much of
the notation used throughout this paper are defined in this section.
We begin with a linear, $2S$-order differential operator $\dop$ that
maps a complex valued function $\psi(x)$, $x\in[0,1]$ to a new
function according to the rule,
\begin{eqnarray}
\dop \psi(x) &=& \sum_{s=0}^{S} \frac{\partial^s}{\partial x^s} \left( a_s(x) \frac{\partial^s
\psi(x)}{\partial x^s} \right) \nonumber \\
&=&a_0(x)\psi(x) + \frac{\partial}{\partial x} \left( a_1(x) \frac{\partial \psi(x)}{\partial x}
\right) + \ldots \nonumber \\ && + \frac{\partial^{S}}{\partial x^{S}} \left( a_{S}
\frac{\partial^{S} \psi(x) }{\partial x^{S}} \right), \label{eq:dopdef}
\end{eqnarray}
where we assume $\psi(x)$ has finite derivatives up to order $2S$.
The coefficients $a_{s}(x)$, $s=0,1,2,\ldots,S$ are finite, real
valued functions on the domain $x \in [0,1]$ with finite derivatives
to order $s$ and satisfy periodic boundary conditions,
\begin{eqnarray}
\frac{\partial^t a_s}{\partial x^t}(0) = \frac{\partial^t a_s}{\partial x^t}(1) & & t=0,1,\ldots,s
\label{eq:coeffcond}
\end{eqnarray}
The minimal smoothness assumed of $a_0(x)$ is continuity on $x \in
[0,1]$. For concreteness, we impose periodic boundary conditions
upon $\psi(x)$ itself,
\begin{eqnarray}
\frac{\partial^t \psi}{\partial x^t}(0) = \frac{\partial^t \psi}{\partial x^t}(1) & &
t=0,1,\ldots,2S \label{eq:boundary}
\end{eqnarray}
although more general homogeneous boundary conditions could be insisted upon. Given the above
definitions, a set of eigenfunctions $\phi_f(x)$ with corresponding real eigenvalues $\lambda_f$ is
defined through,
\begin{equation}
\dop \phi_f(x) = \lambda_f \phi_f(x), \label{eq:eigdef}
\end{equation}
and we order the eigenvalues $\lambda_f$, $f=1,2,3,\ldots$ in
ascending order $\lambda_1 \leq \lambda_2 \leq \lambda_2 \leq
\ldots$. The definition and boundary conditions in Eqs.
\ref{eq:dopdef}-\ref{eq:boundary} guarantee a Hermitian $\dop$,
meaning $\int_0^1dx(\phi_f\dop\phi_{f'}-\phi_{f'}\dop\phi_f)=0$ for
any pair of eigenfunctions $\phi_f,\phi_{f'}$. All the usual
eigenvalue/eigenfunction properties of Hermitian operators follow.
Our task is to estimate a low order ($f=O(1)$) eigenvalue
$\lambda_f$.

The most useful expression of the eigenvalue is the Rayleigh quotient,
\begin{equation}
\lambda_f = \int_0^1 dx \phi_f^*(x) \dop \phi_f(x) = \int_0^1 dx \phi_f^*(x) \lop \phi_f(x)
\label{eq:Rayleigh}
\end{equation}
where we impose unity $L_2$ norm on the eigenfunctions,
\begin{equation}
\| \phi_{f} \|_{L_2} = \left( \int_0^1 dx \phi_f^*(x) \phi_f(x)
\right)^{1/2}
\end{equation}
in anticipation of the quantum algorithm and the operator $\lop$,
derived from $\dop$ by simple integration by parts, is a more
convenient (bilinear) operator to work with due to its symmetric
form,
\begin{equation}
\varphi^*(x) \lop \psi(x) = \sum_{s=0}^{S} \frac{\partial^s\varphi^*(x)}{\partial x^s} a_s(x)
\frac{\partial^s \psi(x)}{\partial x^s} \label{eq:lopdef}
\end{equation}
for any two functions $\varphi(x)$ and $\psi(x)$.

It is useful to work not only in the ``space'' domain $x\in[0,1]$, but also in the ``reciprocal
space'' domain of integers, $k \in \mathbb{Z}$. The connection between the two representations is
defined by the Fourier transforms,
\begin{eqnarray}
\widetilde{\psi}_{k} &=& \int_0^1 dx \exp{(-2\pi i k x)} \psi(x), \nonumber \\
\psi(x) &=& \sum_{k=-\infty}^{\infty} \exp{(2\pi i k x)}
\widetilde{\psi}_{k},
\end{eqnarray}
where tilde will indicate a reciprocal space representation throughout the paper. Our eigenvalue
Eq. \ref{eq:eigdef} is Fourier transformed to
\begin{equation}
\sum_{k'=-\infty}^{\infty} \tlop_{k,k'} \tphi_{f,k'} = \lambda_f \tphi_{f,k}, \label{eq:eigKdef} \\
\end{equation}
where,
\begin{equation}
\tlop_{k,k'} = \sum_{s=0}^{S} (2 \pi i k)^s \ta_{s,k-k'} (2 \pi i k')^s
\end{equation}
is the reciprocal space matrix representation of the operator $\lop$ (and $\dop$). The Rayleigh
quotient of Eq. \ref{eq:Rayleigh} is Fourier transformed to,
\begin{equation}
\lambda_f = \sum_{k,k'=-\infty}^{\infty} \tphi_{f,k} \tlop_{k,k'} \tphi_{f,k'},
\label{eq:Rayleighone}
\end{equation}
where we now have the Euclidean normalization,
\begin{equation}
\left\| \tphi_{f} \right\|_2 = \left( \sum_{k=-\infty}^{\infty} \widetilde{\phi}_{f,k}^*
\widetilde{\phi}_{f,k} \right)^{1/2} = 1,
\end{equation}
consistent with $\| \phi_{f} \|_{L_2}=1$ and our Fourier transform
definition.

In a classical digital computer, discretization of the domain
$x\in[0,1]$ is required so that values $x$ can be represented with a
finite number of bits. For the quantum algorithm we'll be
discussing, discretization of the domain will also be required so
that values $x$ can be identified with a finite number of qubits. We
can then sample the spatial domain at the points
$x=0,1/N,2/N,\ldots,(N-1)/N$, where $N=2^n$ requires $n$ qubits. It
will be more convenient to work with the integers
$\bx=Nx=0,1,2,\ldots,N-1$. The discrete spatial domain of $N$ points
allows us to approximate a function $\psi(x)$ by a vector,
\begin{eqnarray}
\psi^{(N)} = \left(\psi^{(N)}_0,\psi^{(N)}_1,\ldots,\psi^{(N)}_{N-1}\right)
\end{eqnarray}
for computational purposes, where we shall impose Euclidean norm $\|
\psi^{(N)}\|_2=1$. In particular, we wish to generate discretized
approximations $\phi_f^{(N)}$ that approach the continuous problem
eigenvector $\phi_f(x)$ such that taking an ever greater number of
discretization points $N$ gives us the limit $\lim_{N \rightarrow
\infty} \sqrt{N} \phi^{(N)}_{f,\bx} = \phi_f(x)$, the factor
$\sqrt{N}$ accounting for Euclidean normalization of the vector
$\phi_f^{(N)}$ and $L_2$ normalization of the function $\phi_f(x)$.
We discuss how we generate $\phi_f^{(N)}$ and how quantify the
quality of our discrete approximations as a function of $N$ further
below.

In addition to having a discrete approximation to functions
$\psi(x)$, we shall require discrete approximations of the
differential operator $\lop$ of Eq. \ref{eq:lopdef} in the form of
an $N \times N$ matrix $\lop^{(N)}$ acting on vectors $\psi^{(N)}$.
Hence, we'll need $N \times N$ finite difference matrices, which we
shall denote $\Delta_{(s)}^{(N)}$, to approximate derivatives
$\partial^s/
\partial x^s$. There is freedom in choosing finite differences to approximate derivatives,
here we (arbitrarily) choose the forward difference for a concrete example,
\begin{eqnarray}
\left( \Delta_{(1)}^{(N)} \psi^{(N)} \right)_{\bx} &=& N \left( \psi^{(N)}_{\bx+1} -
\psi^{(N)}_{\bx} \right), \label{eq:diffdef}
\end{eqnarray}
and higher order finite differences can be generated by $\Delta_{(s)}^{(N)} = \left(
\Delta_{(1)}^{(N)} \right)^{s}$ for integer $s$. From the very definition of derivatives, we have
$\lim_{N \rightarrow \infty} \sqrt{N}(\Delta_{(s)}^{(N)} \psi^{(N)})_{\bx} = \partial^s \psi(x) /
\partial x^s$ if $\lim_{N \rightarrow \infty} \sqrt{N} \psi^{(N)}_{\bx} = \psi(x)$, the factor $\sqrt{N}$ again accounting for
Euclidean normalization of the vector $\psi^{(N)}$ and $L_2$
normalization of the function $\psi(x)$. The subscript arithmetic
$\bx \pm 1$ in the definition of finite differences is to be
performed modulo-$N$, consistent with the boundary conditions of
Eqs. \ref{eq:coeffcond},\ref{eq:boundary}. The resulting matrix
operator $\lop^{(N)}$ is,
\begin{eqnarray}
\lop^{(N)} = \sum_{s=0}^{S} \left( \Delta^{(N)}_{(s)} \right)^T \cdot \mathrm{Diag}(a^{(N)}_{s})
\cdot \Delta^{(N)}_{(s)}. \label{eq:lopNdef}
\end{eqnarray}
where $(\cdot)^T$ indicates matrix transpose and $\mathrm{Diag}(\cdot)$ indicates a diagonal matrix
with the vector argument along the diagonal. With the above construction for $\lop^{(N)}$, we can
now pose a Hermitian matrix eigenvalue problem,
\begin{eqnarray}
\sum_{\bx'=0}^{N-1} \lop^{(N)}_{\bx,\bx'} \phi^{(N)}_{f,\bx'} = \lambda^{(N)}_{f}
\phi^{(N)}_{f,\bx} \label{eq:eigNdef}
\end{eqnarray}
whose solutions will have the desired properties $\lim_{N
\rightarrow \infty} \sqrt{N} \phi^{(N)}_{f,\bx} = \phi_f(x)$ and
$\lim_{N\rightarrow \infty} \lambda^{(N)}_f = \lambda_f$, with the
obvious restriction $f \leq N$. A reciprocal space description is
useful, for which we introduce the discrete Fourier transforms,
\begin{eqnarray}
\widetilde{\psi}^{(N)}_{k} &=& \frac{1}{\sqrt{N}} \sum_{\bx=0}^{N-1} \omega^{-k \bx} \psi^{(N)}_{\bx}, \nonumber \\
\psi^{(N)}_{\bx} &=& \frac{1}{\sqrt{N}} \sum_{k=-N/2}^{N/2-1} \omega^{ k \bx}
\widetilde{\psi}^{(N)}_{k},
\end{eqnarray}
where $\omega = \exp(2 \pi i / N)$ and reciprocal space has been
truncated to the set $\mathcal{N} = \{ k \in \mathbb{Z} : -N/2 \leq
k \leq N/2+1 \}$. The eigenvectors are assigned unit Euclidean norm
in both $\bx$ and $k$ space representations,
\begin{eqnarray}
\left\| \phi^{(N)}_f \right\|_2 &=& \left( \sum_{\bx=0}^{N-1} \phi^{*(N)}_{f,\bx}
\phi^{(N)}_{f,\bx}
\right)^{1/2} \nonumber \\
&=& \left( \sum_{k \in \mathcal{N}} \phi_{f,k}^{*(N)} \phi_{f,k}^{(N)} \right)^{1/2} = 1,
\end{eqnarray}
so that the discrete analogs of Eqs. \ref{eq:Rayleigh}, \ref{eq:Rayleighone} are
\begin{eqnarray}
\lambda_f^{(N)} &=& \sum_{\bx, \bx' = 0}^{N-1} \phi_{f,\bx'}^{*(N)}\lop^{(N)}_{\bx',\bx}
\phi_{f,\bx}^{(N)} \nonumber \\
&=& \sum_{k,k' \in \mathcal{N}} \tphi_{f,k'}^{*(N)}\tlop^{(N)}_{k',k} \tphi_{f,k}^{(N)}
\label{eq:Rayleightwo}
\end{eqnarray}
which we shall find useful below.

We shall call $|\lambda^{(N)}_f-\lambda_f|$ the \emph{truncation} error, alluding to the fact that
we wish to approximate $\lambda_f$ with $\lambda^{(N)}_f$ while truncating reciprocal space from
all integers $\mathbb{Z}$ to the subset $\mathcal{N}$. We now proceed to show the well known fact
that replacing derivatives by finite differences ultimately limits the convergence of
$\lambda_f^{(N)}$ to $\lambda_f$ as the number of sampling points $N$ increases. Straightforward
application of previously stated definitions gives,
\begin{eqnarray}
\left( \widetilde{ \Delta_{(1)}^{(N)} \psi^{(N)} } \right)_k &=& N \left( \exp \left( 2 \pi i k / N
\right) - 1 \right)\widetilde{\psi}^{(N)}_k \nonumber \\
&=& 2 \pi i k \widetilde{\psi}^{(N)}_k \left( 1 + \Theta \left(\frac{k^2}{N^2} \right) \right)
\label{eq:differror}
\end{eqnarray}
where we have made use of series expansions and the fact that $|k| \leq N/2$ to arrive at the
contribution $\Theta(k^2/N^2)$. The result holds for higher order derivatives.

An important parameter in characterizing truncation error is a reciprocal space cut-off
$k(\phi_f)$, which can be defined for every $\phi_f$. There will always exist a number $k(\phi_f)$
such that $|\widetilde{\phi}_{f,k}|^2 = O(k^{-(4S + 1 + \epsilon)} )$ for all $|k|> k(\phi_f)$ and
some infinitesimal $\epsilon$. This follows simply because $\phi_f$ must be differentiable up to
order $2S$, and therefore the series $\sum_k (2 \pi k)^{4S} |\widetilde{\phi}_{f,k}|^2$ giving the
norm of the $2S^{\mathrm{th}}$ derivative of $\phi_f$ must converge. The eigenvalue spectrum of the
continuous domain operator $\lop$ is unbounded, and it can be shown that $\sup\{k(\phi_f)\}$ does
not exist. However, since we restrict ourselves to $f=O(1)$, we can specify a finite $k(\phi_f)$
independent of $N$. For $N/2>k(\phi_f)$, a reciprocal space cut-off $k(\phi_f^{(N)})$ must also
exist since $\lim_{N \rightarrow \infty} \tphi_{f,k}^{N} = \tphi_{f,k}$. From here on, we shall not
distinguish between $k(\phi_f^{(N)})$ and $k(\phi_f)$ as the precise value of the reciprocal space
cut-off is not needed, but simply its existence. We thus define another subset of reciprocal space
$\mathcal{M} = \{ k \in \mathbb{Z} : |k| < k(\phi_f) \}$.

We have now collected enough ingredients to find the truncation error $|\lambda^{(N)}_f -
\lambda_f|$. We assume that $N/2 > k(\phi_f)$, so that a ``reasonable'' representation of $\phi_f$
can be made on the discretized domain. By ``reasonable'', we mean the eigenvalue $\lambda_f$ can be
estimated using Eq. \ref{eq:Rayleighone} and the truncated reciprocal space $\mathcal{N}$ to give,
\begin{equation}
\lambda_f = \sum_{k,k' \in \mathcal{N} } \tphi^*_{f,k'} \tlop_{k',k} \tphi_{f,k}  +
O\left(N^{-(2S+\epsilon)}\right) \label{eq:truetruncate}
\end{equation}
where the above result arises from the least convergent (highest order derivative) contribution to
$\lambda_f$ in the region $k,k' \notin \mathcal{N}$,
\begin{eqnarray}
 \sum_{k,k' \notin \mathcal{N}} \tphi^{*}_{f,k'} (2 \pi i k)^S \ta_{S,k-k'} (2\pi i k')^S \tphi_{f,k} \nonumber \\
 = \sum_{k,k' \notin \mathcal{N}} O\left(k'^{-(S+1/2+\epsilon)}k^{-(S+1/2+\epsilon)}\right) \nonumber \\
 = O\left(N^{-(2S+\epsilon)}\right)
\end{eqnarray}
where we have made use of $|\widetilde{\phi}_{f,k}| = O(k^{-(2S + 1/2 + \epsilon)} )$ for
$k>k(\phi_f)$. Thus, for $N/2 > k(\phi_f)$, truncation of the reciprocal space sum in Eq.
\ref{eq:Rayleighone} gives $O(N^{-(2S+\epsilon)})$ error.

Using the finite difference error of Eq. \ref{eq:differror}, the reciprocal space matrix
$\tlop^{(N)}$ can be written,
\begin{equation}
\tlop^{(N)}_{k,k'} = \sum_{s=0}^{S} \left[ (2 \pi i k)^s \ta_{s,k-k'} (2 \pi i k')^s \left( 1 +
\Theta \left( \frac{k^2+k'^2}{N^2} \right) \right) \right],
\end{equation}
where we have used the fact that there is some freedom in
approximating $a_s(x)$ by $a^{(N)}_{s,\bx}$. We choose to match
spectral components $\ta^{(N)}_{s,k} = \ta_{s,k}$, and accept that
$a^{(N)}_{s,\bx}$ may exhibit oscillation artifacts (Gibb's
phenomenon) due to discarding the contributions $\ta_{s,k}$ for $k
\in \mathbb{Z}-\mathcal{N}$. Note that the smoothness of $a_s(x)$,
meaning continuity and finite $s$ order derivatives for $x \in
[0,1]$, implies the existence of reciprocal space cut-offs $k(a_s)$.
We use Eq. \ref{eq:Rayleightwo} to decompose,
\begin{eqnarray}
\lambda^{(N)}_f = \sum_{k,k' \in \mathcal{M}} \tphi^{*(N)}_{f,k'} \tlop_{k',k} \tphi^{(N)}_{f,k}\left( 1 + \Theta \left( \frac{k(\phi_f)^2}{N^2} \right) \right)\nonumber \\
+ \sum_{k,k' \in (\mathcal{N}-\mathcal{M})} \tphi^{*(N)}_{f,k'} \tlop_{k',k}
\tphi^{(N)}_{f,k}\left( 1 + \Theta\left( \frac{k^2+k'^2}{N^2} \right) \right) \label{eq:Ntruncate}
\end{eqnarray}
where the error summed over $\mathcal{N}-\mathcal{M}$ is of order,
\begin{eqnarray}
\sum_{k,k' \in (\mathcal{N}-\mathcal{M})}  \Theta \left( \frac{1}{N^2} \right) O \left( \frac{k^2 + k'^2}{k'^{S+1/2+\epsilon} k^{S+1/2+\epsilon} } \right) \nonumber \\
= \Theta \left( \frac{1}{N^2} \right) O\left(\frac{1}{ {k(\phi_f)^{2S-3}}} \right)
\label{eq:Ntruncate2}
\end{eqnarray}
The diminishing contribution of the region $\mathcal{N}-\mathcal{M}$ to $\lambda_{f}^{(N)}$ ensures
that the relative finite difference error $\Theta(k^2/N^2)$ does not approach unity but remains
$\Theta(1/N^2)$. Collecting the results of Eqs. \ref{eq:truetruncate}, \ref{eq:Ntruncate},
\ref{eq:Ntruncate2}, we can express $\lambda_f^{(N)}-\lambda_f$ as,
\begin{widetext}
\begin{eqnarray}
\lambda_f^{(N)} - \lambda_f = \sum_{k,k' \in \mathcal{N} } \tphi^{*(N)}_{f,k'} \tlop_{k',k}
\tphi^{(N)}_{f,k} \left( 1 + \Theta \left( \frac{1}{N^2} \right) \right)- \sum_{k,k' \in
\mathcal{N} } \tphi^{*}_{f,k'} \tlop_{k',k} \tphi_{f,k} \label{eq:truncerrorsum}
\end{eqnarray}
\end{widetext}
where we have dropped the dependence upon $k(\phi_f)$ as it shall be of no further use. We note
that $\tlop^{(N)}-\tlop = \Theta(1/N^2)$ in the reciprocal space $\mathcal{M}$, so we can consider
$1/N^2$ a parameter of expansion in perturbation theory. The lowest order perturbation gives $\|
\delta \tphi_{f}^{(N)} \|_2 = \| \tphi^{(N)}_{f} - \tphi_f \|_2=O(1/N^2)$ for a non-degenerate
$\phi_f$. Degenerate eigenvectors might be perturbed substantially, but this is merely the result
of there being no preferred basis for the span of the degenerate eigenvectors. The same bounds on
truncation error can be shown to apply to the degenerate case. Noting that Eq.
\ref{eq:truncerrorsum} is second order in eigenvector and $\delta \tphi_f$ is orthogonal to
$\tphi_f$, the contribution of $\delta \tphi_f^{(N)}$ to the eigenvalue error is $O(1/N^4)$ and can
therefore be ignored. The relative truncation error is,
\begin{eqnarray}
\left|\frac{\lambda^{(N)}_f-\lambda_f}{\lambda_f} \right| = \Theta\left(\frac{1}{N^2}\right)
\end{eqnarray}
which is the final result of this section. We emphasize that truncation error arises solely from
the uniform discretization of the domain $x\in[0,1]$.

\section{Quantum Algorithm - One Dimension}

\label{sec:quantalg}

We present now the quantum algorithm as it applies to the Hermitian, one-dimensional boundary value
problem discussed in the previous section. We will show the various computational steps, and the
rationale behind them.

First, we set forth some preliminaries. We will represent a vector $\psi^{(N)}$ with a quantum
state composed of $n= \log_2 N$ qubits whose probability amplitudes are encoded as follows,
\begin{equation}
| \psi^{(N)} \rangle = \frac{1}{\sqrt{N}} \sum_{\bx=0}^{N-1} \psi^{(N)}_{\bx} | \bx \rangle,
\end{equation}
where $| \bx \rangle$ is an $n$ qubit state storing the binary representation of $\bx$. Similarly,
the finite difference matrix $\lop^{(N)}$ is mapped to an operator,
\begin{equation}
\lam^{(N)} = \sum_{\bx,\bx'=0}^{N-1} | \bx \rangle \lop^{(N)}_{\bx,\bx'} \langle \bx' |.
\end{equation}
and we define the unitary exponential,
\begin{equation}
U = \exp(i \lam^{(N)} \tau ) = \sum_{q=0}^{\infty}\frac{(i \lam^{(N)} \tau )^q}{q!}.
\end{equation}
where $\tau$ is a dimensionless constant whose value is chosen in advance of the simulation and
where the unitarity of $U$ follows from the Hermitian nature of $\lam^{(N)}$. The constant $\tau$
must be carefully chosen to arrive at a desired accuracy in eigenvalue $\lambda_f^{(N)}$ without an
unnecessarily large number of operations. The prescription for choosing $\tau$ is described further
below in section \ref{sec:compcost}. Note that $\tau$ is now an abstract scaling parameter rather
than the time step of a quantum simulation.

We shall call the register of $n = \log_2 N $ qubits the \emph{accumulator} register. In addition,
a register of $m = \log_2 M$ qubits will be required to count phase steps, which we shall call the
\emph{index} register. Several ancilla qubits will be required, their number depending on the
desired precision for the coefficients $a_s^{(N)}$ that specify $\lam$. The first steps are to load
an initial state $\psi^{(N)}\rangle$ into the accumulator and to form an equal superposition of all
index qubit states, giving a complete state,
\begin{equation}
| \Psi \rangle = \frac{1}{\sqrt{M}} \sum_{j=0}^{M-1}|\psi^{(N)} \rangle|j\rangle,
\end{equation}
The state $|\psi^{(N)}\rangle$ is an initial estimate of the field eigenvector of interest. To
determine the required computational work to arrive at a suitable initial estimate
$|\psi^{(N)}\rangle$, it is useful to decompose the accumulator state in terms of the initially
unknown eigenstates $|\phi^{(N)}_f\rangle$,
\begin{eqnarray}
| \Psi \rangle  = \frac{1}{\sqrt{M}} \sum_{j=0}^{M-1}\sum_{x=0}^{N-1} \alpha_f |\phi_f^{(N)}\rangle
|j\rangle = \sum_{f=0}^{N-1} \alpha_f | \Psi_f \rangle, \label{eq:expandpsi}
\end{eqnarray}
where $\alpha_f = \langle \phi_f^{(N)} | \psi^{(N)} \rangle$. As will be shown, the probability the
Abrams-Lloyd algorithm will give an estimate of eigenvalue $\lambda_f$ in a single iteration is $|
\alpha_f |^2$. To obtain an estimate of $\lambda_f$ with probability approaching unity,
approximately $1/|\alpha_f|^2$ iterations will be required. It is thus necessary for $|\psi^{(N)}
\rangle$ to have a large overlap with $| \phi^{(N)}_f \rangle$ in order to avoid numerous
iterations of the algorithm. The best technique proposed thus far is that of Jaksch and
Papageorgiou \cite{jaksch03}, where a more coarsely defined $\phi_f^{(N_0)}$ is determined first
(ie. $N_0 < N$). According to the analysis of the previous section, a coarse approximation limited
by truncation error will allow one to achieve,
\begin{equation}
|\alpha_f|^2 \leq \left\| \tphi_f - \tphi_f^{(N_0)}\ \right\|_2^2 = \left\| \delta \tphi^{(N_0)}_f
\right\|_2^2 = 1 - O(1/N_0^2)
\end{equation}
for $f=O(1)$. Thus, one might solve for a desired $\phi_f^{(N_0)}$ classically (with cost that we
will discuss later), and load the state $| \phi^{(N_0)} \rangle$ into the accumulator with $\Theta(
N_0 )$ operations.

We shall now follow the linear portion of the algorithm as it operates on a particular component
$|\Psi_f\rangle$, reintroducing the full superposition over all $f$ in the final (nonlinear)
measurement step. The next stage of the algorithm is to apply the unitary $U$ to the accumulator
conditional upon the index to produce the superposition,
\begin{eqnarray}
|\Psi'_f \rangle & = & \frac{1}{\sqrt{M}} \sum_{j=0}^{M-1}{U}^j|\phi^{(N)}_f\rangle |j\rangle \nonumber \\
& = & \frac{1}{\sqrt{M}} \sum_{j=0}^{M-1}\exp(ij\lambda_f^{(N)}\tau)|\phi^{(N)}_f\rangle|j\rangle.
\label{eq:state}
\end{eqnarray}
Only $M$ conditional applications of $U$ are in fact required to form $|\Psi_f'\rangle$ from
$|\Psi_f\rangle$. One applies $U$ conditional on $j>1$, then one applies $U$ conditional on $j>2$
and so forth until the $(M-1)^{\mathrm{th}}$ conditional $U$ is applied for $j=M-1$. The
conditional applications of $U$ can be performed with a single additional ancilla qubit as follows.
With at most $\log M$ logical operations, one can entangle the index register with an ancilla to
form the state $\sum_j |j\rangle|C_{j,j'}\rangle$ where the ancilla $C_{j,j'}=1$ for $j \geq j'$
and $C_{j,j'}=0$ otherwise. The $j'^{\mathrm{th}}$ application of $U$ can be implemented as a $U$
conditional on the ancilla $C_{j,j'}$. The ancilla is then disentangled from the quantum register
by running the initial entangling operation once again.

The operator $U$ acts in the full $n$ qubit Hilbert space of $|\psi^{(N)}\rangle$, which will in
general be prohibitively large, but it is nevertheless possible to efficiently generate an
approximation to $U$ using operations in a few qubit Hilbert space. The structure of $\lam^{(N)}$
is a band diagonal matrix resulting from local operations, and thus it has a block diagonal
representation \emph{in the qubit basis} of the accumulator.

To illustrate explicitly some of the key features of the algorithm at work, it's useful to consider
the simple example where $\dop = \partial / \partial x \{ a(x) (\partial /\ \partial x ) \}$. The
following decomposition is appropriate,
\begin{widetext}
\begin{eqnarray}
\begin{array} {rccc}
\lam^{(N)} = & \underbrace{ N^2\left( \begin{array} {ccccccc}
d_0& & & & & & \\
 &d_1& & & & & \\
 & &d_2& & & & \\
 & & &d_3& & & \\
 & & & &\ddots& & \\
 & & & & & & \\
\end{array} \right)}
- & \underbrace{ N^2\left( \begin{array} {ccccccc}
0&a_1& & & & & \\
a_1&0& & & & & \\
 & &0&a_3& & & \\
 & &a_3&0& & & \\
 & & & &\ddots& & \\
 & & & & & &\\
\end{array} \right)}
- & \underbrace{ N^2\left( \begin{array} {ccccccc}
0& & & & & &a_0 \\
 &0&a_2& & & & \\
 &a_2&0& & & & \\
 & & &0&a_4& & \\
 & & &a_4&0& & \\
a_0& & & & &\ddots& \\
\end{array} \right)} \\
 & \lam^{(N,1)} & \lam^{(N,2)} & \lam^{(N,3)}
\end{array} \label{eq:bigdecomp}
\end{eqnarray}
\end{widetext}
where $d_{\bx}=a_{\bx}+a_{\bx+1}$. The operators $\lam^{(N,p)}$ can be written more compactly,
\begin{eqnarray}
\lam^{(N,1)} &=& N^2 \sum_{\bx} d_{\bx}|\bx\rangle \langle \bx| \nonumber \\
\lam^{(N,2)} &=& - N^2 \sum_{\bx ~ \mathrm{even}} a_{\bx+1} \left\{ |\bx\rangle \langle
\bx+1|+|\bx+1\rangle \langle \bx| \right\} \nonumber \\
\lam^{(N,3)} &=& - N^2\sum_{\bx ~ \mathrm{odd}} a_{\bx+1} \left\{ |\bx\rangle \langle
\bx+1|+|\bx+1\rangle \langle \bx| \right\}, \nonumber
\end{eqnarray}
where $\lam^{(N,1)}$ is diagonal, and $\lam^{(N,2)}$,$\lam^{(N,3)}$ act in one qubit subspaces
(conditional upon $\lfloor \bx / 2\rfloor$) in lieu of the full Hilbert space of $\lam^{(N)}$.

\begin{figure}
\includegraphics[scale=0.4]{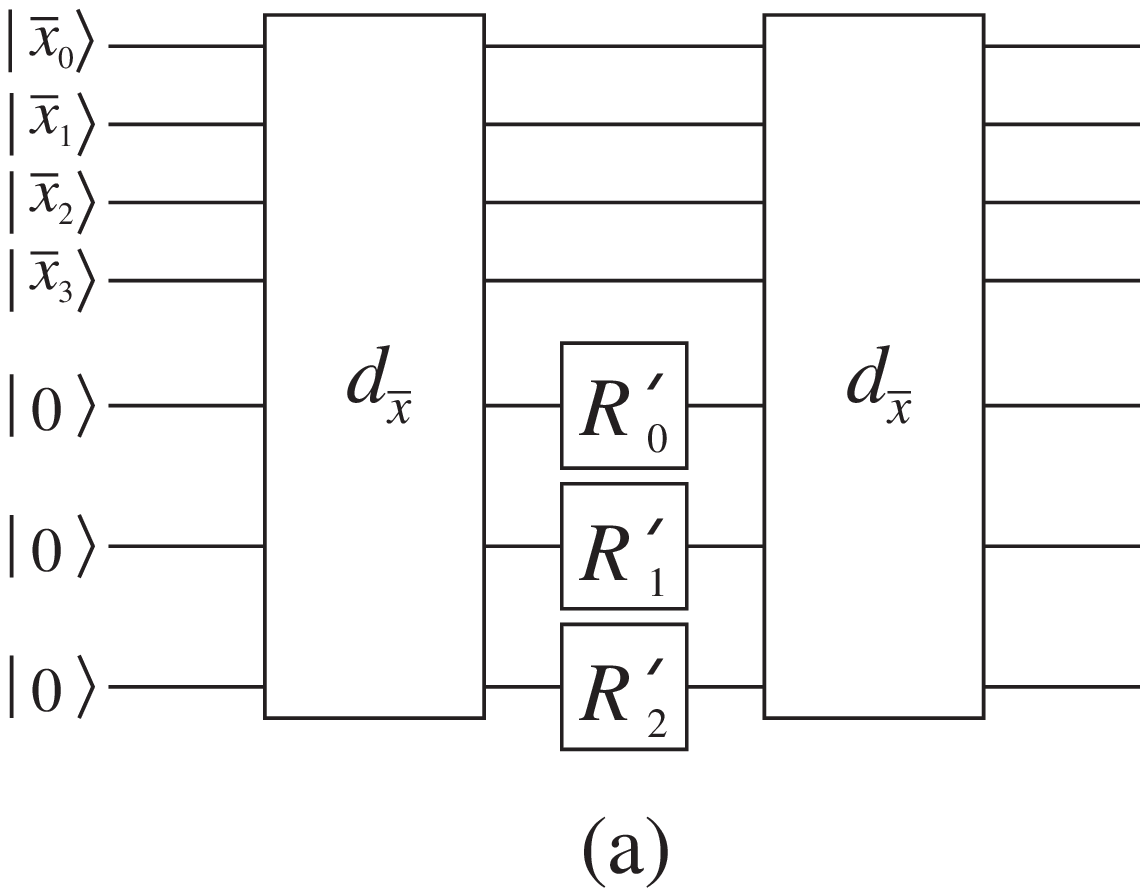}
\vskip 0.5cm
\includegraphics[scale=0.4]{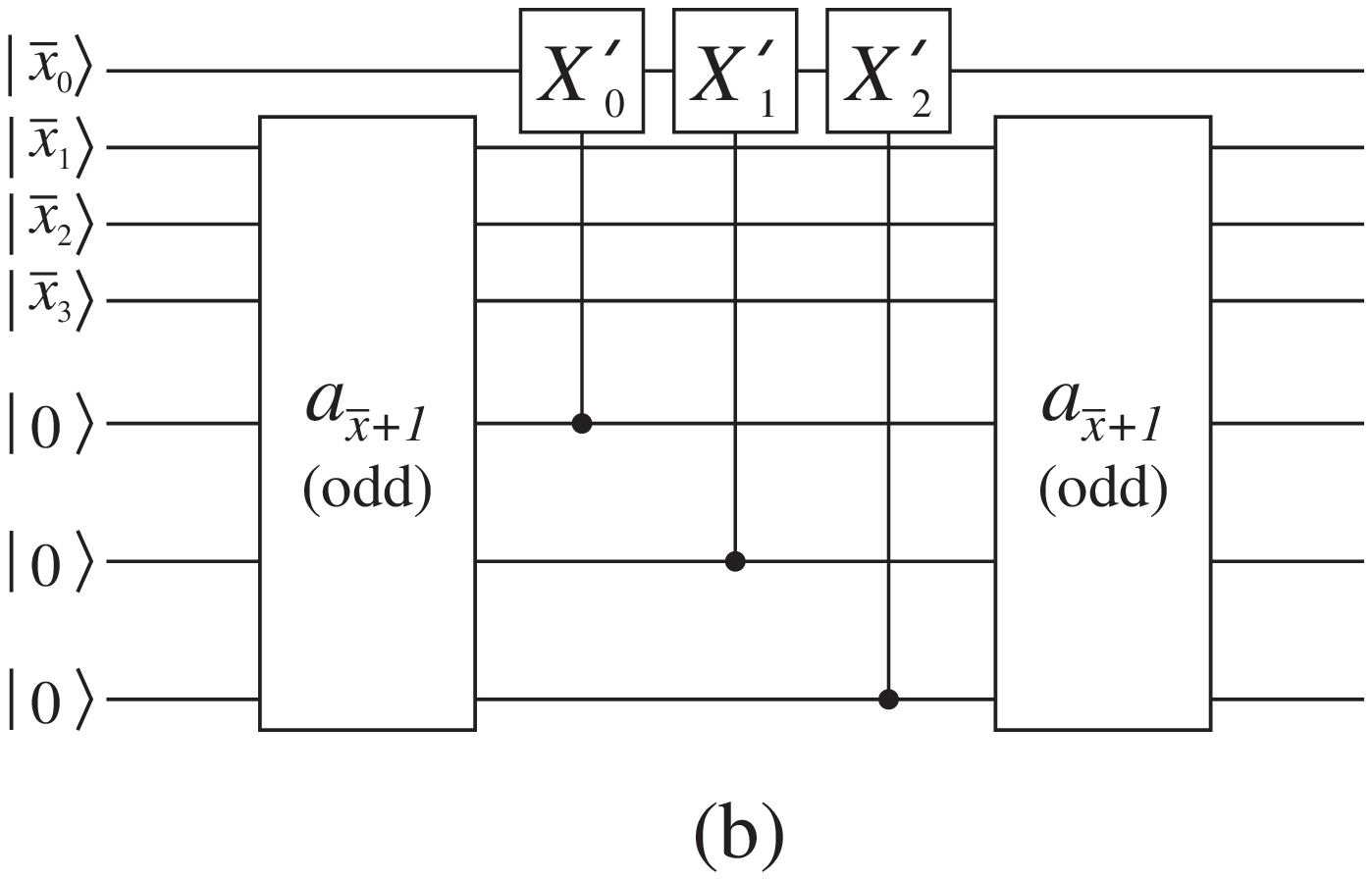}
\vskip 0.5cm
\includegraphics[scale=0.4]{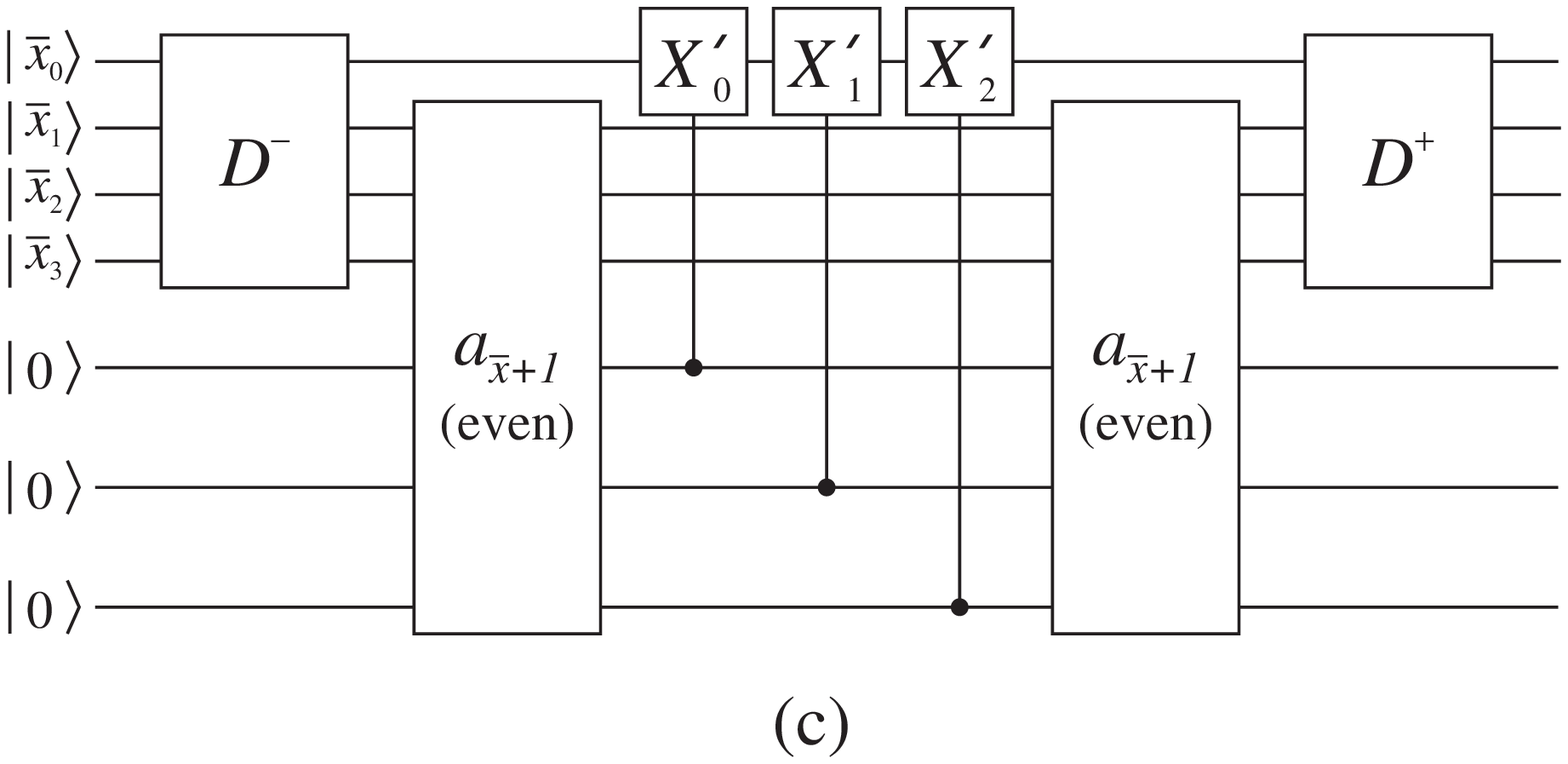}
\caption{\label{fig:circuit} The quantum circuits for applying: (a) $\exp(i\lam^{(N,1)}\tau/2)$,
(b) $\exp(i\lam^{(N,2)}\tau)$, and (c) $\exp(i\lam^{(N,3)}\tau/2)$, to the accumulator qubits
$|\bx\rangle=|\bx_3 \bx_2 \bx_1 \bx_0\rangle$ for the decomposition of Eq. (\ref{eq:bigdecomp})
with $N=2^4$. The ancillae initialized to state $|0\rangle$ are used to store the coefficients
$d_{\bx}$ or $a_{\bx+1}$, to three bit precision with resolution $\delta$: $d_{\bx} = (
\mathbf{d}_{\bx,2}2^2+\mathbf{d}_{\bx,1}2^1+\mathbf{d}_{\bx,0}2^0)\times\delta$ with
$\mathbf{d}_{\bx,j}\in \{ 0,1\}$, and a similar binary description for $a_{\bx+1}$. The circuit
``$d_{\bx}$'' maps $|\bx\rangle |\mathrm{anc}\rangle \rightarrow |\bx\rangle |\mathbf{d}_{\bx}
\oplus \mathrm{anc}\rangle$, while ``$a_{\bx+1}$(even/odd)'' maps $|\bx\rangle |\mathrm{anc}\rangle
\rightarrow |\bx\rangle |\mathbf{a}_{\bx+1}\oplus\mathrm{anc}\rangle$ for even/odd $\bx$. Single
qubit rotations $R'_p = |0\rangle \langle 0 | + \exp(i (N^2 \delta \tau/2) 2^p )|1\rangle \langle
1|$ give the desired diagonal phase shifts for $\exp(i\lam^{(N,1)}\tau/2)$. Single qubit rotations
$X'_p = \exp(i (-N^2 \delta \tau/2) 2^p \sigma_x)$, where $\sigma_x = |0\rangle \langle 1 | +
|1\rangle \langle 0 |$, to implement the desired off diagonal couplings of
$\exp(i\lam^{(N,2)}\tau/2)$ and $\exp(i\lam^{(N,3)}\tau/2)$. The values of $\tau$ and $\delta$ can
be inferred from the restriction that the operator \emph{splitting} error is comparable to
truncation error, described in section \ref{sec:compcost}. The parity shift operators, defined
$D^{\pm}|\bx\rangle = |\bx\pm1\rangle$, are required to shift the block structure of $\lam^{(N,3)}$
so that only operations on the least significant qubit $\bx_0$ need be performed. The $D^{\pm}$ can
be implemented using quantum Fourier transforms (at cost of $O(\log^2 N)$ operations) and single
qubit rotations. Final disentanglement of ancillae is achieved by a second application of
``$d_{\bx}$'' or ``$a_{\bx+1}$''.}
\end{figure}

The unitary $U$ can be approximated to take advantage of the above decomposition in several ways.
For a general decomposition,
\begin{equation}
\lam^{(N)}=\sum_{p=1}^{R}\lam^{(N,p)},
\end{equation}
where for our simple example $R=3$, the Baker-Campbell-Hausdorff formulae can be used to show,
\begin{widetext}
\begin{eqnarray}
U_{\Pi} &=& \prod_{p=1}^{R} \exp\left(i\lam^{(N,p)}\tau/2\right) \prod_{p=R}^{1}
\exp\left(i\lam^{(N,p)}\tau/2\right) \nonumber \\ &=& \exp\left( i \lam^{(N)}\tau -
\frac{i}{3!}\sum_{p,q=1}^{R} \left[ \lam^{(N,p)},\left[\lam^{(N,q)},\lam^{(N,R)} \right]\right]
\tau^3 + O\left( \left\| \lam^{(N)} \right\|_2^4 \tau^4 \right) \right)\label{eq:symproduct}
\end{eqnarray}
\end{widetext}
where terms bilinear in $\lam^{(N,p)}\tau$ are suppressed by the symmetry of the product formula
shown. One may approximate $U$ by $U_{\Pi}$ to take advantage of the efficient implementation of
$\exp(i\lam^{(N,p)}\tau)$ at the cost of introducing error. The quantum circuit for implementing
$U_{\Pi}$ for our simple example $\dop = \partial / \partial x \{ a(x) (\partial /\ \partial x )
\}$ is shown in Fig. \ref{fig:circuit}. The reason for the ease of implementing $U_{\Pi}$ is
apparent in Fig. \ref{fig:circuit}, one applies single qubit unitaries conditional upon the
evaluation of $a_{\bx}$. The Solovay-Kitaev theorem guarantees that the single qubit unitaries can
be implemented to an accuracy $\Theta(1/N^2)$ with $\Theta(\log^c{N})$ universal quantum gates
\cite{nielsen}. We also assume that evaluation of $a_{\bx}$ requires $O(\log N)$ operations.
Roughly speaking, the differentiability of $a(x)$ rules out pathological functions that have
greater complexity.

Approximating $U$ by $U_{\Pi}$ implies that the algorithm will give an estimate of the eigenvalue
$\lambda_{f,\Pi}^{(N)}$ of the operator,
\begin{equation}
\lam^{(N)}_{\Pi} = \lam^{(N)}  + O\left( \left\| \lam^{(N)} \right\|_2^3 \tau^2\right)
\label{eq:lampidef}
\end{equation}
instead of the desired eigenvalue $\lambda_f^{(N)}$. We call the error introduced by using
$U_{\Pi}$ the \emph{splitting} error, which has value $O\left( \left\| \lam^{(N)} \right\|_2^3
\tau^2 \right)$ provided $\|\Lambda^{(N)}\|_2 \tau < 1$. The splitting error will be shown in the
next section to limit the computational efficiency of estimating eigenvalues.

Applying $U_{\Pi}$ rather than $U$, Eq. (\ref{eq:state}) takes the form
\begin{equation}
|\Psi'_{f,\Pi} \rangle  =  \frac{1}{\sqrt{M}}
\sum_{j=0}^{M-1}\exp(ij\lambda_{f,\Pi}^{(N)}\tau)|\phi^{(N)}_{f,\Pi}\rangle|j\rangle.
\label{eq:statepi}
\end{equation}
The eigenvalue is encoded in the phase periodicity of $|\phi_{f,\Pi}\rangle |j\rangle$, and can be
determined to at most the $\pm\pi/M$ precision allowed by a $\log_2 M$ bit representation of a full
$2\pi$ radians. We briefly review the procedure for retrieving the phase \cite{cleve98}, beginning
with the application of the quantum Fourier transform,
\begin{equation}
\mathrm{QFT} = \frac{1}{\sqrt{M}} \sum_{l=0}^{M-1} \sum_{m=0}^{M-1} \exp\left(-2\pi i l m
/M\right)|l\rangle\langle m|,
\end{equation}
to the index qubits. The resulting state is,
\begin{equation}
\mathrm{QFT} | \Psi'_{f,\Pi} \rangle = \sum_{l=0}^{M-1}b_{l,f}| \phi_{f,\Pi}^{(N)}
\rangle|l\rangle.
\end{equation}
with the coefficients,
\begin{equation}
b_{l,f} = \frac{1}{M}\sum_{j=0}^{M-1} \exp\left(i j \left(\lambda_{f,\Pi}^{(N)} \tau - 2 \pi l /
M\right) \right),
\end{equation}
which have square modulus,
\begin{equation}
|b_{l,f}|^2 = \frac{\sin^2 \left[ M\left(2\pi l/M-\lambda _{f,\Pi}^{(N)}\tau\right)/2\right] }{ M^2
\sin^2 \left[ \left(2\pi l/M-\lambda _{f,\Pi}^{(N)}\tau\right)/2\right]}.
\end{equation}
A projective measurement of the index produces $|l'\rangle$ where $|\lambda_{f,\Pi}^{(N)}\tau/2\pi
- l'/M|<1/2M$ with a probability $|b_{l',f}|^2\geq(M^2\sin^2(\pi/2M))^{-1}$. All eigenvalues will
satisfy $|\lambda_{f,\Pi}^{(N)}|<\pi/\tau$ since we will impose $\| \lam^{(N)} \|_2 \tau \ll 1$ (to
be made precise in the next section), so identification of $l'$ will determine an eigenvalue
$\lambda_{f,\Pi}^{(N)}$ uniquely to a precision $\pm \pi/M\tau$.

Since we began not with the desired state alone, but with a superposition
$|\Psi\rangle=\sum_{k=0}^{N-1} \alpha_f | \Psi_f \rangle$, measurement of the index will determine
a particular $\lambda_{f,\Pi}^{(N)}$ with relative probability $|\alpha_f|^2$. It is the initial
trial wavefunction $|\psi^{(N_0)}\rangle$ that determines the probability $|\alpha_f|^2$ of the
eigenvalue/eigenvector pair being selected by a projective measurement.

Upon completion of the eigenvalue readout (via index bits $l'$), the accumulator is left in the
eigenstate $|\phi^{(N)}_f\rangle$. This is useful since it allows further information to be
extracted. For instance, one can efficiently test whether $|\phi_f^{(N)}\rangle$ has a particular
symmetry, such as inversion symmetry about a particular point $\bx$ in the domain. This can serve
as a partial check as to whether the desired $|\phi^{(N)}_f\rangle$ was indeed selected by the
projective measurement.

\section{Computational Cost - One Dimension}

\label{sec:compcost}

We now analyze the computational cost for implementing the Abrams-Lloyd algorithm for the one
dimensional Hermitian problem described in the preceding sections. As pointed out, there are three
sources of error that must be considered to determine the required number of operations for a given
accuracy in eigenvalue estimation.

First, uniform discretization of the continuous problem to $N=2^n$ points on the spatial domain
introduces a truncation error,
\begin{equation}
\left| \frac{\lambda_f^{(N)}-\lambda_f}{\lambda_f} \right| = \Theta \left( \frac{1}{N^2} \right)
\end{equation}
The truncation error quantifies the accuracy with which the discrete problem represents the
continuous problem for low order (ie. $f=O(1)$) eigenfunctions $\phi_f$. To compare algorithms,
classical or quantum, we may ask how many operations are required to achieve the $\Theta(1/N^2)$
accuracy in the solution of the discrete eigenvalue problem.

Second, splitting $\lam^{(N)}$ into parts so as to approximate $U$ with a product $U_{\Pi}$ of
local operators results in what we have termed splitting error. From Eq. \ref{eq:lampidef} the
eigenvalue $\lambda_{f,\Pi}^{(N)}$ of $U_{\Pi}$ is,
\begin{eqnarray}
\lambda_{f,\Pi}^{(N)} &=& \lambda_{f}^{(N)} + O\left( \left\| \lam^{(N)} \right\|_2^3 \tau^2
\right)
\end{eqnarray}
where we choose $\tau$ such that $\|\lam^{(N)}\|_2\tau=\|\lop^{(N)}\|_2\tau<1$. However, from the
the finite difference formula Eq. \ref{eq:diffdef} and the form of $\lop^{(N)}$ in Eq.
\ref{eq:lopNdef}, the spectral radius $\|\lop^{(N)}\|_2 = \Theta(N^{2S})$. Hence, the splitting
error is,
\begin{eqnarray}
\lambda_{f,\Pi}^{(N)} &=& \lambda_{f}^{(N)} + O\left( N^{6S} \tau^2 \right)
\end{eqnarray}
which, unlike truncation error, \emph{increases} polynomially with an increase in the number of
discretization points $N$. The splitting error results from the fact that the product $U_{\Pi}$
creates deviations from the true advancement in phase at high spatial frequencies. For example, in
the system described in Eq. \ref{eq:bigdecomp}, it is the non-commuting nature of advancing even
pairings of points and odd pairings of points that generates an error with spatial frequency
$~N/2$.

Third, the measurement of phase $\lambda_{f,\Pi}^{(N)}\tau$ via the quantum Fourier transform is
limited by the uniform discretization of $2\pi$ radians into $M=2^m$ intervals. The limited phase
resolution allows us to specify $\lambda_{f,\Pi}^{(N)}$ upon completion of the algorithm to a
precision $2 \pi / M \tau$.

The three sources of error allow us to determine the optimal number of index bits $m=\log_2M$, the
value of the constant $\tau$, and thus the complexity of the algorithm. Obviously, there is nothing
gained in solving the discretized problem to an accuracy greater than the truncation error
$\Theta(1/N^2)$ if the goal is to study the continuous problem. We can thus allow the splitting
error $O(N^{6S}\tau^2)$ to be of the same order as the truncation error,
\begin{eqnarray}
\Theta \left(\frac{1}{N^2}\right) \geq O\left( N^{6S} \tau^2 \right) ~~ \rightarrow ~~ \tau \leq
\Omega \left( \frac{1}{N^{3S+1}} \right) \label{eq:taulimit}
\end{eqnarray}
Since $\lambda^{(N)}_{f,\Pi}=O(1)$ for our low order eigenvalue with $f=O(1)$, the phase
advancement $\lambda^{(N)}_{f,\Pi}\tau\leq \Omega(1/N^{3S+1})$ for the low order eigenfunction
becomes exceedingly small. In order to resolve this phase so that our final eigenvalue uncertainty
does not exceed the truncation error, we require
\begin{eqnarray}
\frac{2 \pi}{M \tau} \leq \Theta \left( \frac{1}{N^2} \right) ~~ \rightarrow ~~ M \geq O \left(
N^{3(S+1)} \right)
\end{eqnarray}
thus prescribing the number $m=\log_2 M$ of index register qubits.

The complexity of the eigenvalue estimation can now be stated. The determination of a suitable
initial guess eigenstate $\phi_f^{(N_0)}$ requires the determination of an eigenvector of an $N_0
\times N_0$ problem. This can be done classically in $\Omega(N_0)$ steps, since each of $N_0$
points in the spatial domain description of $\lop^{(N_0)}$ must contribute to the eigenvalue. Near
optimal classical methods are in fact known. In the case of a tridiagonal $\lop^{(N_0)}$, bisection
gives an eigenvalue to $\Theta(1/N_0^2)$ precision with $\Theta(N_0 \log N_0)$ operations
\cite{demmel}. Low order eigenvalues of wider bandwidth $\lop^{(N_0)}$ matrices can be determined
to the same precision with the same order of operations using more complex classical techniques
\ref{note:classicaltechniques}. Only a modest $N_0$ is required for the probability of a successful
iteration of the quantum algorithm, $1-O(1/N_0^2)$, to be comparable to unity. Following the
construction of an initial eigenstate estimate, this estimate must be loaded into the accumulator
register, which can be done in $\Theta(N_0)$ steps. We suppose that $N$ will exceed $N_0$ by a
substantial factor, so that the initial state preparation is a negligible cost compared to the
remainder of the algorithm. The majority of the computational steps in the quantum algorithm are
accounted for by the $M \geq O \left( N^{3(S+1)} \right)$ applications of $U_{\Pi}$, each of which
requires $O(\log^c N)$ gate operations for some constant $c=O(1)$. The final quantum Fourier
transform requires $\Theta(\log^2 M)$ gate operations, a negligible $\log(N)$ contribution compared
to the $M$ applications of $U_{\Pi}$.

Thus, to achieve $\Theta(1/N^2)$ accuracy in the final eigenvalue, at least $O(N^{3(S+1)}\log^c N)$
operations and $\Theta(D\log N)$ qubits are required. In contrast, an eigenvalue can be found using
classical techniques to $\Theta(1/N^2)$ accuracy using $\Theta(N \log N)$ operations. The quantum
algorithm requires significantly more work than classical algorithms for the one dimensional
problem. Nonetheless, we show in the next section that the quantum algorithm is easily extended to
higher dimensional problems where increased efficiency over classical techniques is indeed
possible.

\section{Higher Dimensional Problems}

\label{sec:generalization}

Here we will generalize the results of the one dimensional problem to the multidimensional problem.
Many of the arguments presented in the earlier sections are not specific to the single dimension
domain, and in many cases we can simply replace scalars with vectors. The continuous problem we
wish to solve involves an operator $\dop$ mapping functions $\psi(\mathbf{x})$, defined over a
$D$-dimensional cubic domain $\mathbf{x} \in \mathcal{S} = [0,1]^{\otimes D}$, to functions $\dop
\psi (\mathbf{x})$. Rather than explicitly writing out the general form of a multidimensional
Hermitian operator $\dop$ analagous to the single dimensional operator of Eq. \ref{eq:dopdef}, we
simply state that $\dop$ must satisfy,
\begin{eqnarray}
\int_0^1 dx_1 \cdots \int_0^1 dx_D \left(\phi^*_f \dop \phi_{f'}-\phi^*_{f'} \dop\phi_f\right)=0
\label{eq:hermitdop}
\end{eqnarray}
for any eigenfunctions $\mathbf{\phi}_f$ satisfying $\dop\mathbf{\phi}_f = \lambda_f
\mathbf{\phi}_f$. We can then define an ``equivalent'' bilinear operator $\lop$ that maps any two
vector functions $\psi^*(\mathbf{x})$ and $\varphi(\mathbf{x})$ to a scalar function
$\psi^*\lop\varphi$. This can be done by using the higher dimensional forms of integration by
parts, which in one dimension allowed us to relate $\dop$ to $\lop$. We exclude ``trivial''
problems that are readily expressed as a tensor product of single dimensional problems, $\lop =
\lop_1 \otimes \lop_2 \otimes \ldots \otimes \lop_D$. We are therefore considering problems whose
structure is instead a sum of tensor product terms,
\begin{eqnarray}
\lop = \sum_{\beta=1}^{B} \lop_{\beta,1} \otimes \lop_{\beta,2} \otimes \ldots \otimes
\lop_{\beta,D} \label{eq:multilop}
\end{eqnarray}
for some constant $B>1$, and where the differential order of each one dimensional
$\lop_{\beta,\alpha}$ is $2S_{\beta,\alpha}$. The differential order of $\lop$ is then $2S =
\max_{\beta} \{ \sum_{\alpha} 2S_{\beta,\alpha} \}$. Of course, we retain the Hermitian property
\begin{eqnarray}
\int_0^1 dx_1 \cdots \int_0^1 dx_D \left(\phi^*_f \lop \phi_{f'}-\phi^*_{f'} \lop\phi_f\right)=0
\label{eq:hermitlop}
\end{eqnarray}
and the associated eigenvalue/eigenvector properties. Normalizing the eigenfunctions allows us to
write,
\begin{eqnarray}
\lambda_f = \int_0^1 dx_1 \cdots \int_0^1 dx_D \phi^*_f \lop \phi_f
\end{eqnarray}
which is simply the Rayleigh quotient.

Discretization proceeds as in the single dimensional case, with each domain coordinate
$x_i\in[0,1]$ discretized to $N$ points. Functions $\psi(\mathbf{x})$ are represented by rank $D$
tensors $\psi^{(N)}_\mathbf{\bx}$, ie. for $D=2$ dimensions $\psi^{(N)}_\mathbf{\bx}$ is a matrix
of numbers, for $D=3$ dimensions $\psi^{(N)}_\mathbf{\bx}$ is a ``cube'' of numbers and so forth.
Partial derivatives are converted to finite differences as in the one dimensional case. The
operator $\lop$ is can thus be discretized to a tensor $\lop^{(N)}_{\mathbf{\bx},\mathbf{\bx'}}$.
The truncation error in the multidimensional problem is,
\begin{eqnarray}
\left| \frac{\lambda_f^{(N)}-\lambda_f}{\lambda_f} \right| = \Theta \left( \frac{1}{N^2} \right)
\end{eqnarray}
which is identical to the one dimensional case because the relative finite difference errors on
each coordinate are $\Theta(1/N^2)$.

The implementation of the Abrams-Lloyd algorithm for the multidimensional problem proceeds in a
completely analagous fashion to the one dimensional case, with the number of accumulator qubits $D
\log_2 N = D n$ so as to represent a volume $V=N^D$. As before, an initial estimate of the desired
eigenvector $\phi^{(N)}_f$ is required. A coarse classical simulation can produce an eigenvector
$\phi^{(N_0)}_f$ with $N_0 < N$. Since truncation error scales as $\Theta(1/N^2)$, the required
value of $N_0$ is such that the probability of a successful iteration of the algorithm,
$1-O(1/N_0^2)$, approaches unity. The computational cost is $\Theta(N_0^D \log N_0)$ classical gate
operations for generating the initial eigenstate and $\Theta(N_0^D)$ gate operations to load the
state into the accumulator.

The heart of the algorithm is the controlled application of the unitary $U=\exp(i\lam^{(N)}\tau)$
where $\lam^{(N)}=\sum |\mathbf{\bx}\rangle \lop^{(N)}_{\mathbf{\bx},\mathbf{\bx'}} \langle
\mathbf{\bx'} |$. As before, $U$ acts within a large Hilbert space, so an approximating operator
$U_{\Pi}$ is applied instead. The operator $U_{\Pi}$ is a sequence of operations acting
conditionally upon a much smaller Hilbert space than the full $D\log_2N$ qubits. We quantify the
size of this Hilbert space now. The multidimensional $\lam^{(N)}$ is no longer represented by a
band diagonal matrix, but has the structure of a sum of tensor products as in Eq.
\ref{eq:multilop},
\begin{equation}
\lam^{(N)} = \sum_{\beta=1}^{B} \lam^{(N)}_{\beta,1} \otimes \lam^{(N)}_{\beta,2} \otimes \ldots
\otimes \lam^{(N)}_{\beta,D}
\end{equation}
The local nature of $\lam^{(N)}$ is quantified by the maximum number of states
$|\mathbf{x'}\rangle$ for which $\langle \mathbf{\bx'}| \lam^{(N)} |\mathbf{\bx}\rangle$ is not
zero (maximizing over all possible $|\mathbf{\bx}\rangle$). This volume, $v$, is the maximum
product of matrix bandwidths,
\begin{eqnarray}
v &=& \max_{\beta} \left\{ (2S_{\beta,1}+1) (2S_{\beta,2}+1) \ldots (2S_{\beta,D}+1) \right\} \nonumber \\
&\leq& \left(1 + \frac{2S}{D} \right)^D
\end{eqnarray}
where we have used the restriction $2S = \max_{\beta} \sum_{\alpha} \{ 2S_{\beta,\alpha} \}$ to
arrive at the bound on $v$. It follows that we can split $\lam^{(N)}=\sum_{p=1}^{R}\lam^{(N,p)}$
where the $\lam^{(N,p)}$ act conditionally upon a Hilbert space of $r = \lceil \log_2 v \rceil$
qubits. The size of this reduced Hilbert space is independent of domain size $N^D$, so that
$\exp(i\lam^{(N,p)}\tau)$ can be applied to the requisite accuracy (polynomial in $1/N$) with only
$\Theta(\log^cN)$ universal gates for some constant $c$. As before, we assume that the function
evaluations required for conditional action upon the $r$-qubit subspace entails at most $O(\log N)$
universal gates. The total number of the split up operators $\lam^{(N,p)}$ is bounded $R \leq
v=(1+2S/D)^D$ independently of the domain size $N^D$. Thus, $U_{\Pi}$ can be applied with $O(\log^c
N)$ work for some $c=O(1)$.

The approximation $U_{\Pi}$ can be composed by using a symmetric product as in Eq.
\ref{eq:symproduct} so that the splitting error is,
\begin{eqnarray}
\lambda_{f,\Pi}^{(N)} &=& \lambda_{f}^{(N)} + O\left( \left\| \lam^{(N)} \right\|_2^3 \tau^2
\right)
\end{eqnarray}
More generally, an approximation of $U$ correct to higher order in $\| \lam^{(N)} \|_2 \tau$ can be
implemented \cite{yoshida90, hatano05}, \cite{sornborger99}. For the sake of generality, we assume
we have a product operator $U_{\Pi}$ correct to order $\| \lam^{(N)} \|^{\nu}_2 \tau^{\nu}$, and
set $\nu=2$ to recover the simple symmetric product results. In practice, one can not take $\nu$
arbitrarily large since the number of terms in $U_{\Pi}$ grows exponentially in $\nu$. The optimal
choice of $\nu$ is that which minimizes the overall computational cost.

Using the fact that $2S$ is the differential order of $\lop$, the splitting error becomes,
\begin{eqnarray}
\lambda_{f,\Pi}^{(N)} &=& \lambda_{f}^{(N)} + O\left( N^{2S(\nu+1)} \tau^{\nu} \right)
\end{eqnarray}
The final phase measurement through a quantum Fourier transform proceeds as in the one dimensional
case, with the same precision of $\pm \pi/(M \tau)$ in determining $\lambda^{(N)}_{f,\Pi}$, where
$m=\log_2 M$ is the number of index qubits. Requiring that the final eigenvalue be determined to
the truncation error limit as in the one dimensional case, the same line of reasoning as in the
previous section leads to,
\begin{eqnarray}
\tau \leq \Omega \left( \frac{1}{N^{2(S(1+1/\nu)+1/\nu)}} \right) \nonumber \\
M \geq \ O\left( N^{2(S+1)(1+1/\nu)} \right)
\end{eqnarray}
The computational cost of the algorithm is dominated by the $M$ applications of $U_{\Pi}$, each
application of $U_{\Pi}$ requiring $O(\log^c N)$ number of operations. The computational cost for
the quantum algorithm is,
\begin{equation}
\aleph_Q = O(M\log^c N) = O(N^{2(S+1)(1+1/\nu)}\log^c N)
\end{equation}
in addition to the cost for finding and loading an eigenstate with coarse discretization $N_0$
along each axis. We assume $O(N_0)$. The number of qubits required by the quantum algorithm is
$\Theta(\log N)$.

We now consider classical costs associated with the multidimensional eigenvalue equation.
Discretization and reduction of the continuous problem to a matrix equation results in a sparse
$N^D \times N^D$ matrix with a number of bands depending on the spatial derivatives and dimensions
in the continuous problem. The most efficient and near optimal classical method requires
\begin{eqnarray}
\aleph_C = O(N^D \log N)
\end{eqnarray}
operations in order to attain a low order eigenvalue with truncation error accuracy
$\Theta(1/N^2)$. The method is near optimal in the classical case since the computational cost per
each of $V=N^D$ points in the domain is merely $O(\log N)$. Any classical method must ``visit''
each point in the simulation domain in order for that point to influence the outcome of the
classical calculation, hence the classical computation cost is $\Omega(N^D)$. Of course, the number
of bits required is $\Theta(N^D)$.

The maximum improvement in computational efficiency provided by the quantum algorithm presented is,
\begin{equation}
\max \left\{ \frac{\aleph_C}{\aleph_Q} \right\} = O \left( \frac{N^{D-2(S+1)(1+1/\nu)}}{\log^{c-1}
N} \right)
\end{equation}
with respect to the best known (near optimal) classical algorithm. From the above, we see that the
domain dimension must satisfy $D > 2(S+1)(1+1/\nu)$ in order to see any improvement using the
Abrams-Lloyd algorithm. In particular, we have $S=1$ for Schr\"{o}dinger's equation and we can
identify $D/3$ as the number of particles in space (3 degrees of freedom per particle, neglecting
spin). A many-body eigenvalue calculation is more efficient than classical simulation for particle
number $D/3>(4/3)(1+1/\nu)$. For the case where $U_{\Pi}$ is a simple symmetric product, $\nu = 2$
and we require $D/3 > 2$ in order to see improved computational efficiency. Higher order
approximations, $\nu > 2$ will result in two (spinless) particle calculations already being done
more efficiently using the Abrams-Lloyd algorithm.

We now discuss the generality of the results for domains other than the simple hypercube
$\mathcal{S} = [0,1]^{\otimes D}$ discretized to $V=N^D$ points. A more complex domain
$\mathcal{S'}$ can be had by deleting regions from $\mathcal{S}$ along planes defined by the
uniform discretization scheme. The computational cost incurred is that required to ensure the
probability amplitudes in $|\psi\rangle$ do not ``spill'' into the deleted regions
$\mathcal{S}-\mathcal{S'}$. This is easily done by circuits such as those in Fig.
\ref{fig:circuit}, wherein quantum gates can be used to determine the conditional application of a
few-qubit operator through out the simulation domain. The computational cost is therefore
proportional to the classical cost of determining whether a point $\bx$ is in or out of the
specified domain $\mathcal{S'}$ subtended by the hypercube $\mathcal{S}$. As an explicit example,
the subcircuit $a_{\bx+1}$ of Fig. \ref{fig:circuit} for applying $\exp(i\lam^{(N,2)}\tau)$ of Eq.
\ref{eq:bigdecomp} can be made to compute $|\bx\rangle |\mathrm{anc}\rangle \rightarrow |\bx\rangle
|\mathrm{anc}\rangle$ for $\bx\in\mathcal{S}-\mathcal{S}'$ and $|\bx\rangle |\mathrm{anc}\rangle
\rightarrow |\bx\rangle |a_{\bx+1}\oplus\mathrm{anc}\rangle$ for $\bx\in\mathcal{S'}$. The effect
of this operation is to conditionally apply $\exp(i\lam^{(N,2)}\tau)$ to those points $\bx \in
\mathcal{S'}$. Clearly, $\mathcal{S'}=\mathcal{S}$ is the simplest domain to consider as there is
no added computational cost, but more complex domains are accessible at only the modest cost of
describing the domain with a Boolean function.

\section{Conclusion}

\label{sec:conclusion}

Our analysis of the Abrams-Lloyd algorithm raises several questions. Firstly, it is natural to ask
what sort of qubit phase rotation accuracy is required during the application of $U_{\Pi}$ to the
initial guess eigenstate. The phase that is applied to qubits by the operator $U_{\Pi}$ during the
computation is of the same order as the phase applied to the highest order eigenvector:
$\lambda^{(N)}_{N,\Pi} \tau$ where the eigenvalue $\lambda^{(N)}_{N,\Pi} = \Theta(N^{2S})$ for a
differential operator of order $2S$ and $\tau = \Omega(1/N^{3S+1})$ for a second order splitting
formula. The magnitude of the phase rotations applied to qubits is therefore $\Omega(1/N^{S+1})$.
The relative accuracy with which the phase must be applied is $\Theta(1/N^2)$ if the final
eigenvalue estimation is to be accurate to the truncation error limit of $\Theta(1/N^2)$. Thus the
absolute accuracy required from single qubit rotations is $\Omega(1/N^{S+3})$, independent of the
number of dimensions $D$. The absolute accuracy is a small quantity for very modest values of
$N=100$ (representing a relative eigenvalue accuracy of $10^{-4}$) with a second order operator
($2S=2$). Angular resolution of $10^{-8}$ in the control of qubits represents a technical feat, but
thankfully the principles of fault tolerant quantum computation \cite{preskill98, gottesman98} can
be applied here to lessen the accuracy requirements for \emph{physical qubit} operations.

Secondly, it is tempting to compare the quantum and classical algorithms for the simulation of
dynamical evolution. The Abrams-Lloyd algorithm simulates the dynamics of the Schr\"{o}dinger
equation $\partial \psi / \partial t = \dop \psi$ for some (possibly fictitious) Hamiltonian
represented by $\dop$, but only limited detail of the dynamics in a quantum simulation are
accessible. The probability amplitudes characterizing a register of $D\log N + \log M$ qubits can
result in at most $D\log N+\log M$ classical bits of information being extracted by measurement (by
the Holevo bound). For instance, in order to obtain the eigenvector coefficients $\phi_f^{(N)}$, at
least $\Theta(N^D / \log N )$ iterations of the algorithm would be required. This is in contrast to
a classical simulation of dynamical evolution where $\Theta(N)$ bits would be required to store a
state at a single dynamical step, and $\Theta(NM)$ bits are required to store the entire evolution
of an initial state over $M$ dynamical steps. We emphasize that the strength of the Abrams-Lloyd
algorithm is not in its ability to provide great detail into dynamical evolution but rather in
extracting useful classical information (such as eigenvalues) from a very compact representation of
that dynamical evolution.

Finally, the analysis of the Abrams-Lloyd algorithm raises the question as to why the eigenvalue
convergence for low dimensional problems (ie. small $D$) is less than that of optimal classical
approaches. Part of the answer lies in the classical theory of matrix eigenvalue calculation. An
important tool for numerical estimation of eigenvalues is the Krylov subspace, which is defined to
be the span of the set $\{\psi, A\psi, A^2\psi, \ldots ,A^{M'-1}\psi \}$ for some initial guess
vector $\psi$, some hopefully small constant $M' < N^D$, and some $N^D \times N^D$ matrix $A$ of
which we seek several low order eigenvalues. The Krylov subspace is spanned by at most $M'$
vectors, rather than the full $N^D$ vector space of $A$, and so projecting onto the Krylov subspace
gives an efficient means of estimating eigenvalues/eigenvectors of $A$. If the matrix whose lowest
eigenvalue is sought is $\lam^{(N)}$, then we might choose $A=(\lam^{(N)}-\mu I )^{-1}$ where $\mu$
is an initial estimate of the eigenvalue sought (the eigenvalues of $\lam^{(N)}$ being simply
related to those of $A$). With $A=(\lam^{(N)}-\mu I )^{-1}$, the vector $A^j\psi$ converges
exponentially towards the eigenvector $\phi_f^{(N)}$ whose eigenvalue minimizes
$|\lambda_f^{(N)}-\mu|$. In contrast, if $A=\exp(i\lam^{(N)})$ as in the Abrams-Lloyd algorithm,
there is no such convergence towards a target eigenvector since the eigenvalues of $A$ are of unit
norm. The unitarity of quantum gates restricts eigenvalues to lie on the unit circle in the complex
plane, which is a poor eigenvalue distribution from the perspective of estimating a target
eigenvalue \cite{saad}. This leads to the question of whether controlled decoherence can be used to
produce non-unitarity evolution to accelerate the selection of a target eigenvector \emph{with a
net reduction in gate operations/delay}.

We thank Chris Anderson, Oscar P. Boykin, Salman Habib, Colin Williams, Seth Lloyd, Joseph F. Traub
and Isaac Chuang for stimulating discussions and useful suggestions. This work was supported by the
Defense Advanced Research Projects Agency and the Defense MicroElectronics Activity.

\end{document}